\documentclass[english,letterpaper,twocolumn,showpacs,prl,aps]{revtex4}
\usepackage{graphicx}
\usepackage{amssymb}

\makeatletter

\large  

\input epsf

\makeatother

\usepackage{babel}
\makeatother
\begin{document}

\title{Fermion condensation around a Coulomb impurity in a Weyl semimetal as a manifestation of the Landau zero-charge problem}

\author{Eugene B. Kolomeisky$^{1}$ and Joseph P. Straley$^{2}$}

\affiliation
{$^{1}$Department of Physics, University of Virginia, P. O. Box 400714,
Charlottesville, Virginia 22904-4714, USA\\
$^{2}$Department of Physics and Astronomy, University of Kentucky,
Lexington, Kentucky 40506-0055, USA}

\begin{abstract}
A Coulomb impurity placed in an undoped Weyl semimetal spontaneously surrounds itself with a cloud of condensed Weyl fermions.  We study this system within the Thomas-Fermi approximation.  We find that the ground-state of the system is electrically neutral and exhibits an experimentally accessible Landau zero-charge effect: the impurity charge is screened out at any finite distance in the limit of vanishing impurity size.  Specifically, we show how in this limit the Thomas-Fermi equation for the electrostatic potential transforms into the Gell-Mann-Low equation for the charge.  
  
\end{abstract}

\pacs{71.27.+a, 03.65.Vf}

\maketitle

Over forty years ago Abrikosov and Beneslavski\u{i} \cite{AB} predicted the existence of semimetals having points in the Brillouin zone where the valence and conduction bands meet with a dispersion law that is linear in the wavevector.  Such systems, nowadays called  Weyl semimetals, are likely to be realized in doped silver chalcogenides $Ag_{2+\delta}Se$ and $Ag_{2+\delta}Te$ \cite{silver}, pyrochlore iridates $\mathcal{A}_{2}Ir_{2}O_{7}$ (where $\mathcal{A}$ is Yttrium or a lanthanide) \cite{pyro}, and in  topological insulator multilayer structures \cite{topo}.  Weyl semimetals make it possible to study in condensed matter laboratory settings various effects that were originally proposed in the quantum electrodynamics (QED) context. In this paper we analyze the screening of a Coulomb impurity in a Weyl semimetal; this is related to the vacuum instability with respect to electron condensation due to the electric field a supercharged nucleus \cite{Migdal}.  Our main conclusion is that the character of screening in the Weyl semimetal resembles Landau's "zero-charge" situation in QED \cite{zero_charge,LL4}.  In contrast to QED, the material parameters of a semimetal make this effect experimentally accessible. We emphasize that the physical mechanism of the zero-charge effect described below is different from that originally discussed in QED \cite {zero_charge,LL4}.  Literal analogs of the Landau zero-charge effect were predicted to exist in Weyl semimetals \cite{AB} and in superfluid $^{3}He-A$ \cite{Volovik}.

To demonstrate both the similarities and differences between the semimetal and QED settings, let us consider an electron excitation characterized by a linear dispersion law $\omega=v_{F} k$, where $v_{F}$ is the Fermi velocity and $k$ is the magnitude of the wave vector.  We ask whether the electron (or hole) can be bound by an attractive  point Coulomb impurity of charge $Ze$. Assuming the electron is localized within a spatial range $L$, its energy can be estimated as
\begin{equation}
\label{estimate}
E(L) \simeq \frac{\hbar v_{F}}{L}-\frac{Ze^{2}}{\kappa L}=\frac{\hbar v_{F}}{L} (1- Z\alpha), ~~\alpha=\frac{e^{2}}{\hbar v_{F}\kappa}
\end{equation}
where $\kappa\gtrsim1$ is the dielectric constant.  Assuming the usual order of magnitude estimate $v_{F}\simeq10^{8}cm/sec$, the "fine-structure" constant $\alpha$ can be estimated as $\alpha\simeq1/\kappa$.  Unlike the QED case, both $Z\alpha < 1$ and $Z\alpha > 1$ regimes are experimentally accessible.  Eq.(\ref{estimate}) seems to imply that there is a critical value of the dimensionless impurity charge (number of "protons") $Z_{c_{1}}\simeq 1/\alpha$ such as for $Z<Z_{c_{1}}$ the system is not bound at all, $L\rightarrow \infty$, while for $Z > Z_{c_{1}}$ the energy (\ref{estimate}) is minimized for infinitely localized electron $L\rightarrow 0$.  The latter corresponds to the atomic collapse in superheavy atoms \cite{Migdal,ZP}.  As in the case of QED, the collapse does not occur since in reality the impurity attraction at small distances deviates from the Coulomb law.  

In QED there exists another (truly) critical charge $Z_{c_{2}}\approx170>Z_{c_{1}}$, for which the total energy of the production of an electron-positron pair vanishes and the vacuum becomes unstable with respect to pair creation;  the positron repelled by the nucleus escapes to infinity while the electron remains near the nucleus \cite{ZP}.   In a semimetal the counterpart of $Z_{c_{2}}$ can be estimated based on the observation that due to the zero band gap there is no threshold to creation of electron-hole pairs.  Indeed, let us consider initially bare impurity of dimensionless charge $Z$ and ask whether the system energy can be lowered by condensing single electron and letting the hole escape to infinity.  The effect of electron condensation can be understood in terms of measurable charge of the system as seen at large distances from the impurity.  If $0<Z<1$, the measurable charge $Z-1$ has a sign opposite to that of the bare impurity, and the system energy in this case is proportional to $(1-Z)^{2}$.  This would be lower than the energy of the bare impurity (proportional to $Z^{2}$) only if $Z>Z_{c_{2}}=1/2$.  We conclude that for physically relevant $Z\gtrsim1$ the condensation of Weyl fermions always takes place;  interparticle interactions must be accounted for when the condensate contains more than one fermion.   

In what follows we will determine the ground-state properties of the Weyl semimetal in the presence of the Coulomb impurity as a function of $Z$ and $\alpha$.  The physical arguments given above suggest that the electrons of the "vacuum" (conduction band) spontaneously condense around the impurity while the holes leave the physical picture; the properties of the electron cloud vary with $Z$ and $\alpha$ and are determined by the interplay of attraction to the impurity (promoting electron condensation), and electron-electron repulsion and kinetic energy cost of localization of the electrons (which limit the electron condensation).  For $Z\gg1$ a large number of electrons is expected to be present.  In this limit we can use Thomas-Fermi (TF) theory \cite{LL3} to compute the properties of the system.  This approach parallels that employed in QED \cite{MVP} to understand the physics of supercharged $Z\gg Z_{c_{2}}$ atoms.  Like its more familiar non-relativistic counterpart \cite{LL3}, the TF theory developed below becomes increasingly accurate as $Z\rightarrow \infty$ and is expected to produce reasonable predictions even at $Z\simeq1$ \cite{Lieb}.  

The energy of the system can be written as a functional of the electron number density $n(\textbf{r})$, which is assumed to be slowly varying with position:
\begin{eqnarray}
\label{energy_functional}
\mathcal{E}[n(\textbf{r})]&=&\frac{3 \hbar v_{F}}{4}\left (\frac{6\pi^{2}}{g}\right )^{1/3}\int n^{4/3}(\textbf{r})dV\nonumber\\
&-&e\int \varphi_{ext}(\textbf{r})n(\textbf{r})dV\nonumber\\
 &+& \frac{e^{2}}{2\kappa}\int\frac{n(\textbf{r})n(\textbf{r}')dVdV'}{|\textbf{r}-\textbf{r}'|}
\end{eqnarray}
The first term gives the kinetic energy of free Fermi-gas with linear dispersion relation,  and $g$ counts the number of Weyl nodes within the first Brillouin zone.  For example, $g=24$ in pyrochlore iridates \cite{pyro} and  $g=2$  in a topological insulator multilayer \cite{topo}.  For simplicity we assume an isotropic dispersion characterized by a single Fermi velocity $v_{F}$. The second term is the interaction of the electrons with the impurity.  The external potential $\varphi_{ext}(\textbf{r})$ is a pseudopotential that represents the perturbation of the semimetal caused by the impurity; even though $\varphi_{ext}$ is not entirely of electrostatic origin we will define $\triangle \varphi_{ext} = -4\pi e n_{ext}/\kappa$.  We assume that the impurity charge density $en_{ext}(\textbf{r})$ is spherically-symmetric and localized within a region of size $a$, so that for $r > a$ the potential $\varphi_{ext}(\textbf{r})$ reduces to a purely Coulombic form $\varphi_{ext}(r)= Ze/\kappa r$ corresponding to a net charge $Ze$ within the impurity region.  The first two terms of the energy functional (\ref{energy_functional}), if applied to the question of binding of a single electron in the presence of point impurity ($a=0$),  lead to an estimate analogous to Eq.(\ref{estimate}).  The last term of the energy functional (\ref{energy_functional}) describes the Coulomb repulsion of the electrons.

When the number of condensed electrons is large, their discreteness can be neglected;  then the electron number is not constrained and adjusts itself to minimize the energy functional (\ref{energy_functional}).  Its variational minimization leads to an integral equation
\begin{equation}
\label{int_eq}
\left (\frac{4\pi n(\textbf{r})}{\lambda}\right )^{1/3}-\frac{\kappa \varphi_{ext}}{e}+\int\frac{n(\textbf{r}')dV'}{|\textbf{r}-\textbf{r}'|}=0
\end{equation}
where 
\begin{equation}
\label{lambdadef}
\lambda = \frac{2g\alpha^{3}}{3\pi}
\end{equation}
is a dimensionless parameter that characterizes the relative strength of electron-electron interactions and zero-point motion represented by the first and third terms in Eqs.(\ref{energy_functional}) and (\ref{int_eq}), respectively.  The weakly-interacting or quantum-dominated regime corresponds to the $\lambda\ll1$ domain while the classical or strongly interacting limit is recovered in the $\lambda\gg1$ case.  A wide range of values for $\lambda$ are experimentally accessible.  Taking in Eq.(\ref{int_eq}) the $r\rightarrow \infty$ limit gives (for $\lambda\neq0$) a relationship
\begin{equation}
\label{n_normalization}
\int n(\textbf{r})dV=Z
\end{equation} 
i.e. the semimetal succeeds in giving complete screening of the impurity charge;  the Coulomb impurity in the Weyl semimetal resembles an artificial atom in vacuum except that it is made of Weyl fermions.  

When the interactions are weak $\lambda \ll1$, the screening is weak, too.  Then neglecting the last term in Eq.(\ref{int_eq}) seems legitimate and we find 
\begin{equation}
\label{density_weak screening}
n(\textbf{r})= \frac{\lambda}{4\pi} \left (\frac{\kappa\varphi_{ext}(\textbf{r})}{e}\right )^{3}
\end{equation}
Inside the impurity region where $\varphi_{ext}\simeq Ze/\kappa a$ we have $n\simeq \lambda Z^{3}/a^{3}$ thus implying that the number of condensed electrons residing at $r<a$, is estimated as $\lambda Z^{3}$.  The latter is much smaller than the total number $Z$ of condensed electrons provided $\lambda Z^{2}\ll1$ (thus tightening the initial $\lambda\ll1$ constraint).  Outside the impurity region Eq.(\ref{density_weak screening}) becomes 
\begin{equation}
\label{outside_n_weak screening}
n(\textbf{r})=\frac{\lambda Z^{3}}{4\pi r^{3}}
\end{equation}
Thus  number of condensed electrons confined between concentric spheres of radii $a$ and $r>a$ is given by
\begin{equation}
\label{asymptotic_electron_number}
N(r> a) = \lambda Z^{3} \ln\frac{r}{a}
\end{equation}
Eqs.(\ref{outside_n_weak screening}) and (\ref{asymptotic_electron_number}) clearly contradict the normalization condition (\ref{n_normalization}) and the resolution is that there exists a spatial scale $R$ beyond which the $\lambda Z^{2}\ll1$ perturbation theory and thus Eqs.(\ref{outside_n_weak screening}) and (\ref{asymptotic_electron_number}) fail.  This scale could be called the cloud size; it can be estimated from the condition $N(R>a)\simeq Z$ with the result $R(\lambda)\simeq a\exp(const/\lambda Z^{2})$ (a more precise expression will be given below).    

Eq.(\ref{int_eq}), like its non-relativistic counterpart \cite{LL3}, can be converted into a differential equation by application of the Laplacian $\triangle$:
\begin{equation}
\label{diff_eq}
\frac{e}{\kappa}\triangle \left (\frac{4\pi n}{\lambda}\right )^{1/3}=-\frac{4\pi e}{\kappa}(n_{ext}-n) 
\end{equation}
Comparing Eq.(\ref{diff_eq}) with the Poisson equation for the full self-consistent electrostatic potential $\varphi$ (that combines the impurity potential $\varphi_{ext}$ and electrostatic potential of the electron cloud) shows that   
\begin{equation}
\label{sc_potential}
\varphi= \frac{e}{\kappa}\left (\frac{4\pi n(\textbf{r})}{\lambda}\right )^{1/3}
\end{equation}
The weak-screening $\lambda Z^{2}\ll1$ analysis conducted above corresponds to the approximation $\varphi=\varphi_{ext}$.  In terms of the physical potential (\ref{sc_potential}) Eq.(\ref{diff_eq}) becomes
\begin{equation}
\label{diff_eq_phi}
\triangle \left (\frac{\kappa\varphi}{e}\right ) =-4\pi n_{ext} +\lambda\left (\frac{\kappa\varphi}{e}\right )^{3}
\end{equation}
For $\kappa=1$ and $g=2$ (spin-1/2 fermions in vacuum) Eq.(\ref{diff_eq_phi}) reduces to the ultra-relativistic limit of the QED TF equation \cite{MVP}. 

It is convenient to replace $\varphi$ with the function $\zeta$
\begin{equation}
\label{def_zeta}
\varphi(r)= \frac{e}{\kappa r}\zeta \left (\frac{r}{a}\right )  
\end{equation}
which, via Gauss's theorem, is related to the dimensionless measurable impurity charge at position $r$ (defined as the total charge inside the sphere of radius $r$) as:
\begin{equation}
\label{charge_connection}
-r^{2}\frac{\partial(\kappa\varphi/e)}{\partial r}=\zeta(x)-x\frac{d\zeta}{dx}\equiv \zeta(l)-\zeta'(l),~~l=\ln\frac{r}{a}
\end{equation} 
Substituting (\ref{def_zeta}) into (\ref{diff_eq}) we obtain the equation
\begin{equation}
\label{diff_eq_zeta_of_x}
\zeta''(l)-\zeta'(l)\equiv x^{2}\frac{d^{2}\zeta}{dx^{2}}=-4\pi x^{3}a^{3}n_{ext}+\lambda\zeta^{3}
\end{equation}
Substituting (\ref{def_zeta}) into (\ref{sc_potential}) we find for the electron density the expression of the form
\begin{equation}
\label{n_via_zeta}
n=\frac{\lambda}{4\pi r^{3}}\zeta^{3} \left (\frac{r}{a}\right ) 
\end{equation}
Since $n$ must decay faster than $1/r^{3}$ (see Eq.(\ref{n_normalization})), the solution to Eq.(\ref{diff_eq_zeta_of_x}) is subject to the boundary condition $\zeta(\infty)=0$ and to the normalization condition implied by Eqs.(\ref{n_via_zeta}) and (\ref{n_normalization}).  

The weak screening $\lambda Z^{2}\ll1$ analysis can be further extended by treating the cubic term of (\ref{diff_eq_zeta_of_x}) perturbatively.  Then the lowest order solution is $\zeta=Z$  outside the impurity region.  The next order gives for $r>a$
\begin{equation}
\label{zeta_weak screening}
\zeta=Z(1-\lambda Z^{2}l)=Z(1-\lambda Z^{2}\ln\frac{r}{a}), ~~~\lambda Z^{2}l\ll1
\end{equation}
We observe that the expression for the measurable charge (\ref{charge_connection}) computed with the help of Eq.(\ref{zeta_weak screening}) agrees with Eq.(\ref{asymptotic_electron_number}) to logarithmic accuracy.   Then to logarithmic accuracy the perturbative expression (\ref{zeta_weak screening}) may be regarded as a charge itself;  like Eq.(\ref{asymptotic_electron_number}) it is applicable if $\lambda Z^{2}l\ll1$ i. e. within the Weyl cloud.  Eq.(\ref{zeta_weak screening}) tells us that even within the cloud, the physical electrostatic potential $\varphi$ and the density of condensed electrons $n$ decrease with $r$ slightly faster than $Ze/\kappa r$ and $\lambda Z^{3}/4\pi r^{3}$, (see Eq.(\ref{outside_n_weak screening})), respectively.  We also observe that Eq.(\ref{zeta_weak screening}) resembles Dirac's QED formula \cite{Dirac} which predicted a logarithmic momentum dependence of the physical electron charge due to the effects of vacuum polarization.  

As the interactions parameterized by $\lambda$ (\ref{lambdadef}) increase, the role of quantum-mechanical effects weakens.  In the classical $\lambda =\infty$ limit the screening problem has a very simple solution:  $n(\textbf{r})=n_{ext}(\textbf{r})$ and $\varphi=0$ (implied by Eqs.(\ref{int_eq}) or (\ref{diff_eq}) and (\ref{sc_potential})).  This means that the screening response of the semimetal to the presence of the impurity is that of an ideal classical conductor:  the system is locally neutral and the electric field is zero everywhere.       

When $n_{ext}(\textbf{r})$ is a smooth function and $\lambda \gg1$,  Eqs.(\ref{diff_eq}) and (\ref{sc_potential}) imply  
\begin{equation}
\label{leading_strong_screening_n}
n(\textbf{r})=n_{ext}(\textbf{r})+\frac{1}{(16\pi^{2}\lambda)^{1/3}}\triangle n_{ext}^{1/3}(\textbf{r})+...
\end{equation}
\begin{equation}
\label{leading_strong_screening_phi}
\varphi(\textbf{r})=\frac{e}{\kappa}\left (\frac{4\pi n_{ext}(\textbf{r})}{\lambda }\right )^{1/3}+...
\end{equation}  
We see that for $\lambda $ large but finite the screening is nearly complete and a small deviation (of order $\lambda^{-1/3}$)  of the electron number density from $n_{ext}(\textbf{r})$ is responsible for the potential given by Eq.(\ref{leading_strong_screening_phi}).  This imbalanced charge is localized in the region of maximal change of the impurity charge distribution $n_{ext}(\textbf{r})$, i.e. near the impurity boundary; the same applies to the residual electric field $-\nabla \varphi$.  These conclusions resemble earlier results \cite{MVP}.

In practice the strong screening solution (\ref{leading_strong_screening_n}) and (\ref{leading_strong_screening_phi}) is only relevant inside of the impurity region because the impurity boundary has atomic width;  a physically appropriate model of the impurity charge distribution is a function $n_{ext}(\textbf{r})$ that is strictly zero everywhere outside the impurity.  For example, for a uniformly charged ball model of the impurity Eq.(\ref{leading_strong_screening_n}) predicts that $n=0$ for $r>a$.  However this can be only true  in the $\lambda=\infty$ case; a better approximation requires a solution of the non-linear equations (\ref{diff_eq}) or (\ref{diff_eq_phi}) that does not rely on the linearization approximation about $n=n_{ext}(\textbf{r})$.  Eq.(\ref{leading_strong_screening_phi}) allows us to estimate the residual charge within the impurity region to be of the order $Ze/(\lambda Z^{2})^{1/3}\ll Ze$, thus implying that most of the condensed electrons are inside while only a small number of the order $Z/(\lambda Z^{2})^{1/3}$ is outside the impurity region.

This leads to a physical picture of screening as follows:

*In the weak-screening limit, $\lambda Z^{2}\ll1$, nearly all condensed electrons are outside of the impurity region;  of the order $Z$ of them reside within the Weyl cloud of size $R(\lambda)\simeq a\exp(const/\lambda Z^{2})$.  Within the cloud, the condensed electron density decays approximately as $1/r^{3}$ as described by Eqs.(\ref{n_via_zeta}) and (\ref{zeta_weak screening}), while the electrostatic potential given by Eqs.(\ref{def_zeta}) and (\ref{zeta_weak screening}) is nearly the same as that of the bare impurity, meaning that the screening charge is having small effect.  Outside of the cloud the electron density and potential decay faster than $1/r^{3}$ and $1/r$, respectively, in a way yet to be determined.  

*As $\lambda Z^{2}$ increases, more electrons invade the impurity region and the size of the cloud shrinks.  This picture is consistent with what we already know about the regime of strong screening $\lambda Z^{2}\gg1$: For $\lambda Z^{2}=\infty$ the screening is complete and all the condensed electrons reside within the impurity region;  the electric field is zero everywhere.  For finite $\lambda Z^{2}\gg1$ the screening is nearly complete; the majority of the condensed electrons reside within the impurity region, while a small minority is found outside.  Based on these results we argue that no qualitative changes occur in the screening response as the system evolves between the regimes of weak and strong screening.  Below we put this assertion on a solid ground by showing that for any $\lambda$ the large-distance screening response is universal (in the strong-screening limit this will include the entire outside region).

For $l=\ln(r/a)\gg1$ we can neglect in Eq.(\ref{diff_eq_zeta_of_x}) the second-order derivative term $\zeta''(l)$ compared to $\zeta'(l)$ (the alternate choice $|\zeta''(l)|\gg |\zeta'(l)|$ leads to an inconsistency).  Then for large $l$ we have $\zeta(l)\propto 1/\sqrt{l}$ and $|\zeta(l)| \gg |\zeta'(l)|$, so that in the $l\gg1$ limit $\zeta(l)$ (see Eq.(\ref{charge_connection})) may be called the enclosed charge.  Well outside the impurity region $l\gg1$ Eq.(\ref{diff_eq_zeta_of_x}) simplifies to 
\begin{equation}
\label{GL}
\frac{d\zeta}{dl}=-\lambda\zeta^{3}
\end{equation}
This equation is mathematically identical to the Gell-Mann-Low equation \cite{GL,LL4} for the physical charge in QED \cite{zero_charge} reflecting the effects of vacuum polarization.  Eq.(\ref{GL}) exhibits the phenomenon of "zero charge" - no matter what the "initial" value of $\zeta$ is, the system "flows" to the zero charge fixed point $\zeta=0$ as $l\rightarrow \infty$, i.e. the impurity charge has been completely screened.  Since $l=\ln(r/a)$, the $l\rightarrow \infty$ limit means either the point impurity limit $a\rightarrow 0$ and arbitrary distance $r$ or fixed impurity size $a$ and $r\rightarrow \infty$.  Eq.(\ref{GL}) can be integrated with the result
\begin{equation}
\label{0_charge_solution}
\zeta^{2}(l\gg1)=\frac{\zeta_{0}^{2}}{1+2\lambda \zeta_{0}^{2}l}
\end{equation}
where $\zeta_{0}$ is an integration constant determined by the net charge within the impurity region.  At sufficiently large length scales, $l\rightarrow \infty$, the initial value $\zeta_{0}$ is unimportant and $\zeta$ is independent of $Z$, for any value of $\lambda$.   

Substituting Eq.(\ref{0_charge_solution}) into the expressions for the potential (\ref{def_zeta}) and electron density (\ref{n_via_zeta}) we find our main results
\begin{equation}
\label{potential_large_r}
\varphi(r)=\frac {\zeta_{0}e}{\kappa r \sqrt {1+ 2\lambda \zeta_{0}^{2}\ln\frac{r}{a}}}\rightarrow\frac{e}{\kappa r\sqrt{2\lambda\ln\frac{r}{a}}}
\end{equation}
\begin{equation}
\label{density_large_r}
n(r)=\frac{\lambda \zeta_{0}^{3}}{4\pi r^{3}(1+2\lambda \zeta_{0}^{2}\ln\frac{r}{a})^{3/2}}\rightarrow\frac{\lambda}{4\pi r^{3}(2\lambda \ln\frac{r}{a})^{3/2}}
\end{equation}
applicable in the $l=\ln(r/a)\gg1$ limit.  These equations can be viewed as interpolation formulas for arbitrary $r\geqslant a$, $\lambda$, and $Z$, with $\zeta_{0}$ having the meaning of the net dimensionless charge within the impurity region.  The logarithmic terms in the denominators of Eqs. (\ref{potential_large_r}) and (\ref{density_large_r}) are relevant at the scales $r$ exceeding
\begin{equation}
\label{cloud_size}
R(\lambda)\simeq a e^{1/2\lambda \zeta_{0}^{2}}
\end{equation}
which is the size of the Weyl cloud.

The total number of the condensed electrons confined between the spheres of radii $a$ and $r>a$ computed with the help of Eq.(\ref{density_large_r})
\begin{equation}
\label{weak_screening_interpolation_N}
N(r>a)\approx \zeta_{0}\left (1-\frac{1}{\sqrt{1+2\lambda \zeta_{0}^{2}\ln\frac{r}{a}}} \right )
\end{equation}
is finite and equal to $\zeta_{0}$ as $r\rightarrow \infty$.  

In the weak screening limit $\lambda Z^{2}\ll1$ when $\zeta_{0}=Z$, the size of the Weyl cloud (\ref{cloud_size}) becomes $R(\lambda) \simeq a \exp(1/2\lambda Z^{2})$, sharpening the previous estimate.  Moreover Eqs.(\ref{0_charge_solution}) and (\ref{weak_screening_interpolation_N}) include the perturbative $\lambda Z^{2}\ln(r/a) \ll 1$ results, Eqs.(\ref{zeta_weak screening}) and (\ref{asymptotic_electron_number}), respectively.  We also observe that in the weak screening limit the Weyl cloud described by Eqs.(\ref{potential_large_r}) and (\ref{density_large_r}) is an extremely weakly localized object:  only about $30\%$ of the condensed electrons are found within the range (\ref{cloud_size}).

In the strong screening limit $\lambda Z^{2}\gg 1$ when $\zeta_{0}\simeq Z/(\lambda Z^{2})^{1/3}\ll Z$, the size of the Weyl cloud (\ref{cloud_size}) approaches the impurity size. This means that asymptotic universal regime described by Eqs.(\ref{potential_large_r}) and (\ref{density_large_r}) is the solution to the strong-screening problem practically everywhere outside the impurity region.  
 
We thank G. E. Volovik for correspondence and D. Vaman and P. Arnold for comments.  This work was supported by US AFOSR Grant No. FA9550-11-1-0297.

\end{document}